\documentclass[a4paper,fleqn,usenatbib]{mnras}

\usepackage{newtxtext,newtxmath}
\usepackage[T1]{fontenc}
\usepackage{ae,aecompl}
\pdfoutput=1

\usepackage{graphicx}	% Including figure files
\usepackage{amsmath}	% Advanced maths commands
\newcommand{\omc}{\Omega_{c}}

\newcommand{\omb}{\Omega_{b}}

\newcommand{\lcdm}{$\Lambda$CDM }
\newcommand{\hii}{H{\footnotesize II} {}}

\title[Cosmological constraints from CMB and new HIIG]{Cosmological Constraints using the newest VLT-KMOS HII Galaxies and the full {\em Planck} CMB spectrum.}

\author[P. Tsiapi et al.]{
Pavlina Tsiapi,$^{1}$\thanks{E-mail: ptsiapi@mail.ntua.gr}
Spyros Basilakos,$^{2,3}$
Manolis Plionis,$^{3,4}$
Roberto Terlevich,$^{5,6}$
\newauthor
Elena Terlevich,$^{5}$
Ana Luisa Gonzalez Moran,$^{5}$
Ricardo Chavez,$^{7}$
Fabio Bresolin,$^{8}$
\newauthor
David Fernandez Arenas,$^{9,5}$
Eduardo Telles$^{10}$
\\
$^{1}$National Technical University of Athens, 9 Iroon Polytechniou St.,
15780, Greece\\
$^{2}$Academy of Athens Research Center for Astronomy \& Applied Mathematics, Soranou Efessiou 4, 11-527 Athens, Greece\\
$^{3}$National Observatory of Athens, P.Pendeli, Athens, Greece\\
$^{4}$\mbox{Physics Dept., Aristotle Univ. of Thessaloniki,Thessaloniki 54124, Greece}\\
$^{5}$Instituto Nacional de Astrof\'{i}sica, Optica y Electr\'{o}nica,Tonantzintla,C.P. 72840, Puebla, M\'{e}xico\\
$^{6}$Institute of Astronomy, University of Cambridge, Cambridge, CB3 0HA, UK\\
$^{7}$CONACyT-Instituto de Radioastronom\'{i}a y Astrof\'{i}sica, UNAM, Campus Morelia, C.P. 58089, Morelia, M\'{e}xico\\
$^{8}$Institute for Astronomy, University of Hawaii, 2680 Woodlawn Drive, 96822 Honolulu,HI USA\\
$^{9}$Kavli Institute for Astronomy and Astrophysics, Peking University, Beijing 100871, China\\
$^{10}$Observatorio Nacional, Rua Jos\'{e} Cristino 77, 20921-400 Rio de Janeiro, Brasil
}

\date{Accepted XXX. Received YYY; in original form ZZZ}

\pubyear{2021}

\begin{document}
\label{firstpage}
\pagerange{\pageref{firstpage}--\pageref{lastpage}}
\maketitle

\begin{abstract}
  We present novel cosmological constraints based on a joint analysis of our \hii galaxies (HIIG) Hubble relation with the full {\em Planck} Cosmic Microwave Background anisotropy spectrum and the Baryon Acoustic Oscillations (BAO) probes. The \hii galaxies span a large redshift range ($0.088\leq z \leq 2.5$), reaching significantly higher redshifts than available SNIa and hence they probe the cosmic expansion at earlier times. Our independent constraints compare well with those based on the ``Pantheon'' compilation of SNIa data, which we also analyse. We find our results to be in agreement with the conformal \lcdm model within $1\sigma$.  We also use our HIIG data to examine the behaviour of the dark energy equation of state parameter under the CPL parameterisation, $w = w_0+w_a z/(1+z)$, and find consistent results with those based on SNIa, although the degeneracy in the parameter space as well as the individual parameter uncertainties, when marginalizing one over the other, are quite large.
\end{abstract}


%%%%%%%%%%%%%%%%%%%%%%%%%%%%%%%%%%%%%%%%%%%%%%%%%%

%%%%%%%%%%%%%%%%% BODY OF PAPER %%%%%%%%%%%%%%%%%%

\section{Introduction}

Since 1998 the scientific community of cosmology has been at a consensus over the expansion of the Universe  \citep{Riess:1998cb,Perlmutter:1998np};
type Ia Supernovae (SNIa) observations have proven beyond doubt that we live in an expanding Universe, whose rate of expansion is accelerating. 
Over the last two decades the combined analysis 
of Cosmic Microwave Background  (CMB)  anisotropies \citep[eg.][]{Jaffe_2001,Pryke_2002,Spergel:2006hy, refId0, Ade:2015xua, Aghanim:2018eyx}
%Jaffe et al 2001, Pryke 2002
with Baryon Acoustic Oscillations (BAOs) \citep[eg.][]{Eisenstein:2005su,Blake2011} and Hubble parameter measurements
\citep[eg.][]{Chavez12,Freedman_2012,Riess:2016jrr, Riess18,FernandezArenas2018}
  have shown that the cosmic fluid appears to be dominated by an unknown component which is usually referred to as dark energy (hereafter DE). From the phenomenological viewpoint, the unknown nature of dark energy is reflected in the equation of state parameter (hereafter EoS), namely $w=p_{\rm D}/\rho_{D}$, where the quantities $\rho_{D}$ and $p_{D}$ are the density and pressure of the DE fluid respectively. It is well known that one of the main targets of observational cosmology is to constrain $w$ as well as to test %for 
its evolution.

In this context, the basic cosmological parameters (including the EoS)  
are constrained by a combination of (a) relatively low redshift ($z\le 2$)
cosmological probes
(SNIa and BAOs: \citealt{Riess:1998cb, Perlmutter:1998np,Hicken_2009, Amanullah_2010, Riess_2011,Suzuki_2012, Betoule:2014frx, Scolnic:2017caz})  
and (b)   high  redshift probes ($z\sim 1000$ {\em Planck}  CMB  fluctuations;  eg., \citealt{refId0,Ade:2015xua,Aghanim:2018eyx}).
Such a combination is essential in order to minimize degeneracies among the cosmological parameters; 
however, if we assume that $w$ is a function of redshift then strong degeneracies persist.
It is important to note that probing
%Obviously, the absence of determinations of cosmological parameters at
intermediate redshifts ($2\le z \le 10$), where the maximum difference in
cosmological models take place \citep[cf.][]{Plionis_2011},
%appear to be the key in constraining $w(z)$.
is a prerequisite for effectively constraining $w(z)$.

\hii galaxies are alternative and effective tracers of the Hubble relation for two reasons: (a) they can be observed up to high redshifts $z \sim 3.5$  \citep{Terlevich:2015toa, Chavez16}, where the distance
modulus is more sensitive to the cosmological parameters, and (b) there is a tight relation between the $H\beta$ luminosity and the emission line velocity dispersion, which provides an effective HIIG distance indicator. Indeed it has been proven 
\citep[cf.][]{Melnick:1999qb, Siegel:2004xs, Plionis_2011, Chavez12,Chavez14,Chavez16,Terlevich:2015toa,Gonzalez-Moran:2019uij,Gonzalez-Moran:2021drc}
that this relation can be used as an alternative cosmological tracer. 
Despite the fact that the scatter of the HIIG distance modulus is larger (by a factor of 2) than that of high-$z$ SNIa, this demerit is fully 
compensated by the fact that HIIG are observed to larger redshifts than SNIa, where, as discussed above, the degeneracies for different DE models are
%largely 
reduced \citep{Plionis_2011}.  It is important to remark that over the last ten years our team has published a series of papers on this subject.

In the current article, we use our new set of high spectral resolution observations of high-z HIIG obtained with VLT-KMOS \citep{Gonzalez-Moran:2021drc} along with published HIIG data \citep{Gonzalez-Moran:2019uij,Terlevich:2015toa}, and combine them with the full {\em Planck} spectrum \citep{Aghanim:2018eyx}, Type 1a Supernovae data from the Pantheon dataset \citep{Scolnic:2017caz}, Baryon Acoustic Oscillations data points \citep{Beutler_2011,Kazin:2014qga,Ross:2014qpa,Alam:2016hwk}, and the value for the $H_0$ parameter reported in \citet{Riess18}  in order to place constraints on the main cosmological parameters, as well as to check for dynamical DE. In this work we consider the \lcdm model as the fiducial model, and also explore the popular Chevallier-Polarski-Linder  \citep{Chevallier:2000qy,Linder:2002et} parameterisation (hereafter CPL).

This paper is organised as follows: 
in section 2 we provide the basic cosmological equations, in section 3 we briefly describe the data samples used in the current analysis. In section 4 we present our analysis along with comments on our results, while in section 5 we draw our conclusions and discuss future research prospects.

\section{Basic cosmological equations}
In the context of general relativity we consider that the universe is a
self-gravitating fluid, endowed with a spatially flat homogeneous and isotropic geometry. Furthermore, we also assume that the cosmic fluid is dominated by non-relativistic matter plus a dark energy component with an equation of state given by $p_{\rm D}=w(a)\rho_{\rm D}$ which is responsible for 
the accelerated expansion of the universe. Within this framework, the Hubble parameter takes the form:
\begin{equation}
E^{2}(a)=\frac{H^{2}(a)}{H_{0}^{2}}= \Omega_{m,0}a^{-3}+\Omega_{\rm D,0}X(a),  \label{nfe1}
\end{equation}
where 
\begin{equation}
X(a)={\rm exp}\left(3\int^{1}_{a} d{\rm lny}[1+w(y)]\right).
\end{equation}
with $E(a)$ denoting the normalized Hubble flow, $H_{0}$ the Hubble constant, $a(t)$ the scale factor and $w(a)$ the equation of state parameter. Notice, that $\Omega_{m,0}$ and $\Omega_{D,0}(\equiv 1-\Omega_{m,0})$ are the fractional matter and dark energy density parameters at the present time, respectively.

For the concordance $\Lambda$CDM model we have $w(a)\equiv - 1$ 
and thus $X(a)=1$. On the other hand we can parameterise the unknown 
form of the EoS parameter by using the CPL 
parameterisation. This approach 
has been largely discussed in the literature, namely
the equation of state parameter is expressed 
as a first order Taylor expansion around the present 
time, $a(t_{0})=1$: 
\begin{equation}
w(a)=w_{0}+w_{a}(1-a), 
\end{equation}
hence
\begin{equation}
X(a)=a^{-3(1+w_{0}+w_{a})}{\rm exp}\left[3w_{a}(a-1)\right],
\end{equation}
where $w_{0}$ and $w_{a}$ are constants. Notice that for $a\rightarrow 0$ 
we have $w\simeq w_{0}+w_{a}$, while prior to the present time ($a=1$)
the EoS parameter tends to $w_{0}$.

Finally, the luminosity distance corresponding to the spatially flat Friedmann-Lema\^{\i}tre-Robertson-Walker metric reads: 
\begin{equation}
d_{L}(z)=c(1+z)\int_{0}^{z} \frac{du}{H(u)},
\label{eq:lumdist}
\end{equation}
and the distance modulus is given by 
\begin{equation}
m-M=5{\rm log}d_{L}+25,
\label{eq:distmod}
\end{equation}
where the distance $d_L$ is given in units of Mpc, $z$ is the redshift and $1+z=a(z)^{-1}$.

 \begin{figure*}
	\includegraphics[width=\textwidth]{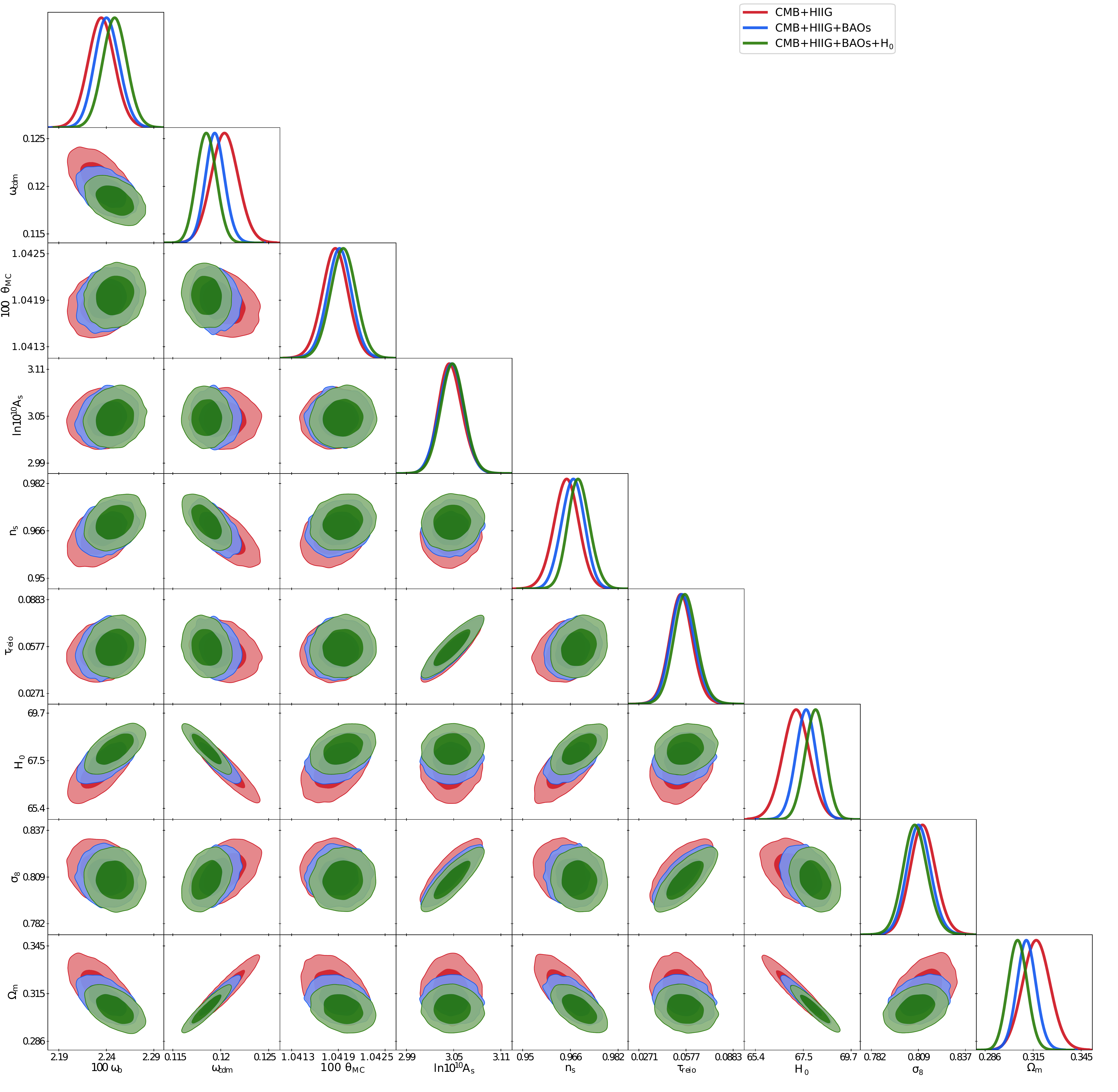}
	\caption{$1\sigma$ and $2\sigma$ contour plots for the \lcdm model using CMB+HIIG data (red), CMB+HIIG+BAO data (blue), and CMB+HIIG+H0 data (green).}
	\label{fig:hiig}
\end{figure*}

 \begin{figure*}
	\includegraphics[width=\textwidth]{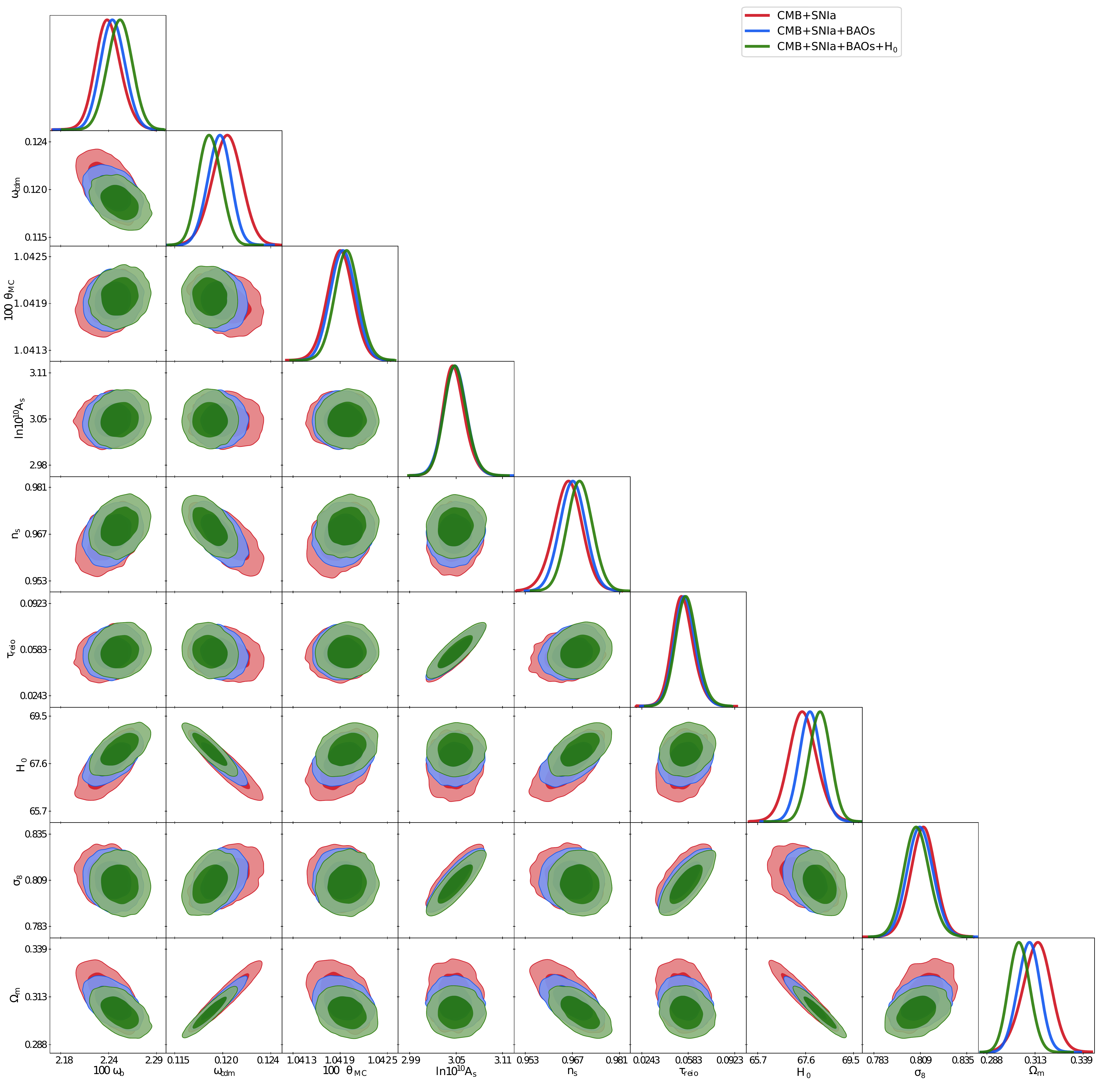}
	\caption{Similar to \ref{fig:hiig}, with SNIa data instead of HIIG.}
	\label{fig:snia}
\end{figure*}

\section{Cosmological Data}
In the current work we use the publicly available {\em Planck} CMB spectrum in combination with \hii galaxies data and other cosmological probes, described below, in order to constrain the standard set of cosmological parameters of the 
$\Lambda$CDM and CPL models, respectively.
Notice that for the former case the parameter space is
$$(\omb h^2, \omc h^2, 100\theta_{MC}, \ln(10^{10}A_s) , n_s , \tau_{\text{reio}})$$
while for the latter we have 
$$(\omb h^2, \omc h^2, 100\theta_{MC}, \ln(10^{10}A_s), n_s, \tau_{\text{reio}}, w_0, w_a),$$
where $\omb$ and $\omc$ are the fractional baryonic and dark matter density parameters at the present time, respectively, $\theta_{MC}$ is the angular size of the sound horizon at recombination, $A_{s}$ is the amplitude of the primordial power spectrum, $n_{s}$ is the spectral index and $\tau_{\text{reio}}$ is the optical depth at reionisation.

In the following we briefly present the type of cosmological data used in our current statistical analysis.
\begin{itemize}
\item \textit{H{\footnotesize \textit{II}} galaxies data}:
In a series of papers \citep{Gonzalez-Moran:2021drc,Gonzalez-Moran:2019uij,FernandezArenas2018,Chavez16,Terlevich:2015toa,Chavez14phd,Plionis:2009wp, Plionis_2011,Melnick:1999qb,Plionis_2009} 
\hii galaxies have been used as alternative tracers of the Hubble relation, hence extending Hubble relation on redshifts beyond the range of available SNIa data. Here we utilize the \hii galaxies dataset discussed in \citet{Gonzalez-Moran:2021drc}.
The current HIIG sample contains 181 objects, which can be separated in the local sample (107 HIIG with $z<0.16$) and a high-z sample based on 29 KMOS, 15 MOSFIRE, 6 XShooter and 24 literature objects (for details see \citealt{Gonzalez-Moran:2021drc}).
The luminosity of the HIIG in the sample is attributed mostly to a recent single starburst ($< 5Myr$).

\item \textit{Supernovae (SNIa)}:
  We also use the binned data from the Pantheon Supernova type Ia (SNIa) dataset \citep{Scolnic:2017caz}, which combines the confirmed SNIa sources discovered by the Pan-STARRS1 (PS1) Medium Deep Survey with a large number of other surveys, to compare the two distinct and independent tracers of the Hubble function.

%We would like to clarify that with the present analysis we do not wish to compete with the SNIa results. Our aim is rather to extend the Hubble relation to as high redshifts as possible.

\item \textit{CMB:} We use the full CMB spectrum from the Planck 2018 data release \citep{Aghanim:2018eyx}. Specifically, we focus on the full (TT+TE+BB+lowE) spectrum, and compare results  with those based on the CMB shift parameter likelihood, considered in \citet{Terlevich:2015toa} and \citet{Gonzalez-Moran:2021drc}. To this end we use the publicly available CLASS software \citep{Blas_2011} within the Monte Python MCMC code \citep{Brinckmann:2018cvx,Audren:2012wb}.

\item \textit{Baryon Acoustic Oscillations}:  We use the BAO probe with measurements from various sources: the WiggleZ Dark Energy Survey \citep{Kazin:2014qga}, the BOSS DR12 survey \citep{Alam:2016hwk}, one single 6DFGS point 
	\citep{Beutler_2011} and the Main Galaxy Sample of Data Release 7 of 
	Sloan Digital SkySurvey (SDSS-MGS) \citep{Ross:2014qpa}.

\item \textit{Hubble constant}: Lastly, we utilize the value 
$H_0 = 73.48 \pm 1.66 \text{\rmfamily km s}^{-1}{}\text{\rmfamily Mpc}^{-1}$
	 as reported in \citet{Riess18}. 
\end{itemize}

\begin{table*}
	\centering
	\caption{The results from the \lcdm analysis using the HIIG data. The 1-$\sigma$ error bars are quoted. The parameter $H_0$ is displayed in $\text{\rmfamily km s}^{-1}{}\text{\rmfamily Mpc}^{-1}$ units.}
	\label{tab:tabhl}
	\begin{tabular}{lrrr} % four columns, alignment for each UPDATED FINAL
		\hline
		Parameters & CMB+HIIG & CMB+HIIG+BAOs & Planck+HIIG+BAOs+$H_0$ \\
		\hline
		$\omc h^2$                & $0.1206 \pm 0.0014$  & $0.1196 \pm 0.0011$  & $0.1188 \pm 0.0011$  \\
		
		$\omb h^2$                & $0.02234 \pm 0.00015$ & $0.02241 \pm 0.00014$ & $0.02249 \pm 0.00014$ \\
		
		$\tau_{\text{reio}}$  & $0.0543^{+0.0038}_{-0.0042}$ & $0.00556^{+0.0037}_{-0.0041}$ & $0.0571^{+0.0038}_{-0.0042}$ \\
		
		$n_s$                   & $0.965 \pm 0.004$    & $0.967 \pm 0.004$  & $0.969 \pm 0.004$ \\

		$\ln(10^{10}A_s)$       & $3.046\pm 0.016 $  & $3.046 \pm 0.016$ & $3.047 \pm 0.017$ \\
		
		$100\theta_{MC}$        & $1.0419 \pm 0.0003 $ & $1.0420 \pm 0.0003$ & $1.0420 \pm 0.0003$ \\
		\hline
		$\sigma_8$              & $0.812 \pm 0.008$  & $0.810 \pm 0.007$  & $0.808 \pm 0.007$ \\
		
		$\Omega_{m}$            & $0.317 \pm 0.008$ & $0.311 \pm 0.006$ & $0.306 \pm 0.006$\\
		
		$H_0$                   & $67.19 \pm 0.61 $   & $67.58 \pm 0.47$ & $68.11 \pm 0.45$\\
		
		\hline
	\end{tabular}
\end{table*}

\section{Observational constraints}
We perform a joint analysis of the \hii galaxies Hubble relation with the full {\em Planck} CMB spectrum and the BAO probes
in order to place constraints on the cosmological parameters of the concordance $\Lambda$CDM and the popular CPL models.
We shall consider that the above datasets can be treated as statistically independent probes of the models, which is a reasonable assumption since the individual probes are based on different cosmic objects which trace mostly different redshift ranges and are based on different physical mechanisms.
Only in one case, where we make combined use of both the HIIGs and the SNIa as tracers of the Hubble expansion, does the latter assumption come under question, given the fact that there is some spatial overlap between the two probes, which in turn could introduce correlations in the statistical analysis.

While this could be important, unfortunately at the moment there is no standard way to account for it, given the lack of the full correlation matrix between the samples. Therefore, following the standard $\chi^{2}$ procedure, we have assumed in all cases that the different  datasets are uncorrelated, which is equivalent to simply summing over their corresponding $\chi^{2}$ functions.

\begin{table*}
	\centering
	\caption{The results from the \lcdm analysis using the Pantheon data. The 1-$\sigma$ error bars are quoted. The parameter $H_0$ is displayed in $\text{\rmfamily km s}^{-1}{}\text{\rmfamily Mpc}^{-1}$ units.}
	\label{tab:tabsl}
	\begin{tabular}{lrrr} % four columns, alignment for each UPDATED FINAL
		\hline
		Parameters & CMB+SNIa & CMB+SNIa+BAOs & Planck+SNIa+BAOs+$H_0$ \\
		\hline
		$\omc h^2$                & $0.1198 \pm 0.0013$   & $0.1192 \pm 0.0010$  & $0.1184 \pm 0.0010$  \\
		
		$\omb h^2$                & $0.02238 \pm 0.00015$ & $0.02243 \pm 0.00014$ & $0.02251 \pm 0.00014$ \\
		
		$\tau_{\text{reio}}$                  & $0.0547^{+0.0038}_{-0.0041}$ & $0.0558^{+0.0036}_{-0.0041}$ & $0.0571^{+0.0038}_{-0.0042}$ \\
		
		$n_s$                   & $0.966 \pm 0.004$  & $0.967 \pm 0.004$  & $0.970 \pm 0.004$ \\
		
		$\ln(10^{10}A_s)$       & $3.045\pm 0.016$     & $3.046 \pm 0.016$ & $3.047 \pm 0.017$\\
		
		$100\theta_{MC}$        & $1.0419 \pm 0.0003$& $1.0420 \pm 0.0003$ & $1.0421 \pm 0.0003$ \\
		
		\hline
		
		$\sigma_8$              & $0.810 \pm 0.008$  & $0.810 \pm 0.007$  & $0.807 \pm 0.007$ \\
		
		$\Omega_m$              & $0.314 \pm 0.008$  & $0.310 \pm 0.006$  & $0.305 \pm 0.006$ \\
		
		$H_0$                   & $67.47 \pm 0.60$   & $67.76 \pm 0.44$   & $68.14 \pm 0.44$ \\
		\hline		
		
	\end{tabular}
\end{table*}

Within this context, for the case of the concordance \lcdm model, we constrain the standard parameter space, namely $(\omb h^2, \omc h^2, 100\theta_{MC}, ln(10^{10}A_s), n_s , \tau_{\text{reio}})$. In Table \ref{tab:tabhl} we provide an overall presentation of the observational constraints imposed by the joint analysis of HIIG, CMB spectrum and BAO probes, while in Table \ref{tab:tabsl} we present the corresponding results based on using SNIa instead of HIIG. Obviously, this is a comparison of the performance of the two independent tracers, HIIG and SNIa, in probing the cosmic expansion  within this particular model. Moreover in figures \ref{fig:hiig} and \ref{fig:snia} we show the $1\sigma$ and $2\sigma$ contours in various parameter planes. 

Inspecting the aforementioned constraints, we verify that the results based on \hii galaxies 
are in excellent agreement with those based on SNIa. It is noteworthy that this agreement extends also to the relative errors associated with the quoted values. This, however, is because the majority of the sampled parameters are mostly constrained by the CMB spectrum.
Note that both joint analyses (based on either HIIG or SNIa) provide results which are consistent (within $1\sigma$) with those provided by the Planck collaboration \citep[see][]{Ade:2015xua,Aghanim:2018eyx}.

Regarding the well known Hubble constant tension problem, ie., the fact that local measurements (ie.   
$H_0 = 73.48 \pm 1.66 \text{\rmfamily km s}^{-1}{}\text{\rmfamily Mpc}^{-1}$ as measured by SNIa \citealt{Riess18})  and HIIG \citep{Chavez12,FernandezArenas2018} are in $\sim 4\sigma$ tension with the $z\sim 1100$ value provided by {\em Planck}, ie., $H_{0}= 67.36 \pm 0.54$ km/s/Mpc \citep{Aghanim:2018eyx}, our results are consistent with the latter measurements, as expected from the fact that the CMB spectrum probe dominates the $H_{0}-$constraints imposed by the joint analysis.
We also note that such results are in agreement with those of \citet{Shanks:2018rka}, who found $H_{0}= 67.6 \pm 1.52$ km/s/Mpc utilizing the GAIA parallax distances of Milky Way Cepheids. 
We need to emphasize however all local Universe studies support larger values of $H_{0}$ \citep[cf.][]{Lima:2012jm, Beaton_2016, FernandezArenas2018,DiValentino:2021izs}.

We now focus on the CPL parameterisation,
which allows a dynamic evolution of the equation of state (EoS) parameter, $w = w_0 + w_a(1-a)$, and hence 
it introduces two more parameters in the original parameter space.

We repeat our statistical analysis but now for the parameter space $(\omb h^2, \omc h^2, 100\theta_{MC}, \ln(10^{10}A_s), n_s, \tau_{\text{reio}}, w_0, w_a)$. Due to the large number of free parameters we restrict our analysis to the maximum data combinations of: CMB+HIIG+BAOs+$H_0$ and CMB+SNIa+BAOs+$H_0$.
However, we also allow the combined use of both Hubble expansion tracers, with the caveats discussed previously, providing
results based on the joint analysis CMB+HIIG+SNIa+BAOs+$H_0$.

\begin{table*}
	\centering
	\caption{The results from the CPL analysis using the HIIG data (second column), Pantheon data (third column), and both datasets (fourth column). 
The 1-$\sigma$ error bars are quoted. The parameter $H_0$ is displayed in $\text{\rmfamily km s}^{-1}{}\text{\rmfamily Mpc}^{-1}$units.}
	\label{tab:tabhsw}
	\begin{tabular}{lrrr} % two columns, alignment for each
		\hline
		Parameters & CMB+HIIG+BAOs+$H_0$ & CMB+SNIa+BAOs+$H_0$ & CMB+HIIG+SNIa+BAOs+$H_0$\\
		\hline
		$\omc h^2$ & $0.1203 \pm 0.0013 $ & $0.1201 \pm 0.0013$&$0.1202\pm 0.0012$ \\
		
		$\omb h^2$ & $0.02237 \pm 0.0015$ & $0.02236 \pm 0.00035$ &$0.02238\pm0.00015$ \\
		
		$\tau_{\text{reio}}$     & $0.0545\pm 0.0078$ & $0.0545\pm 0.0076$ &$0.0544\pm 0.0075$ \\
		
		$n_s$      & $0.965 \pm 0.005$ & $0.965\pm 0.004$ &$0.965\pm 0.004$ \\
		
		$\ln(10^{10}A_s)$ & $3.046 \pm 0.020$ & $3.046\pm 0.019$ & $3.046\pm 0.017$ \\
		
		$100\theta_{MC}$  &$1.0420 \pm 0.00031$ & $1.0419 \pm 0,00030$& $1.0419 \pm 0.00031$ \\
		
		$w_0$ & $-1.06 \pm 0.26$ & $-1.02 \pm 0.08$ & $-1.012 \pm 0.07$ \\
		
		$w_a$  & $ -0.092 \pm 0.45 $ & $ - 0.19 \pm 0.32 $ & $ -0.29\pm 0.31$ \\
		
		\hline
		
		$\sigma_8$        &$0.829 \pm 0.022$ & $0.830\pm 0.020$ & $0.829 \pm 0.019$ \\
		
		$\Omega_m$       & $0.293 \pm 0.021$ & $0.299 \pm 0.015 $ & $0.301 \pm 0.014$\\
		
		$H_0$             & $69.92 \pm 0.89$ & $69.27 \pm 0.71 $ & $69.20 \pm 0.70$  \\
		\hline
	\end{tabular}
\end{table*}

\begin{figure}
	\includegraphics[width=\columnwidth]{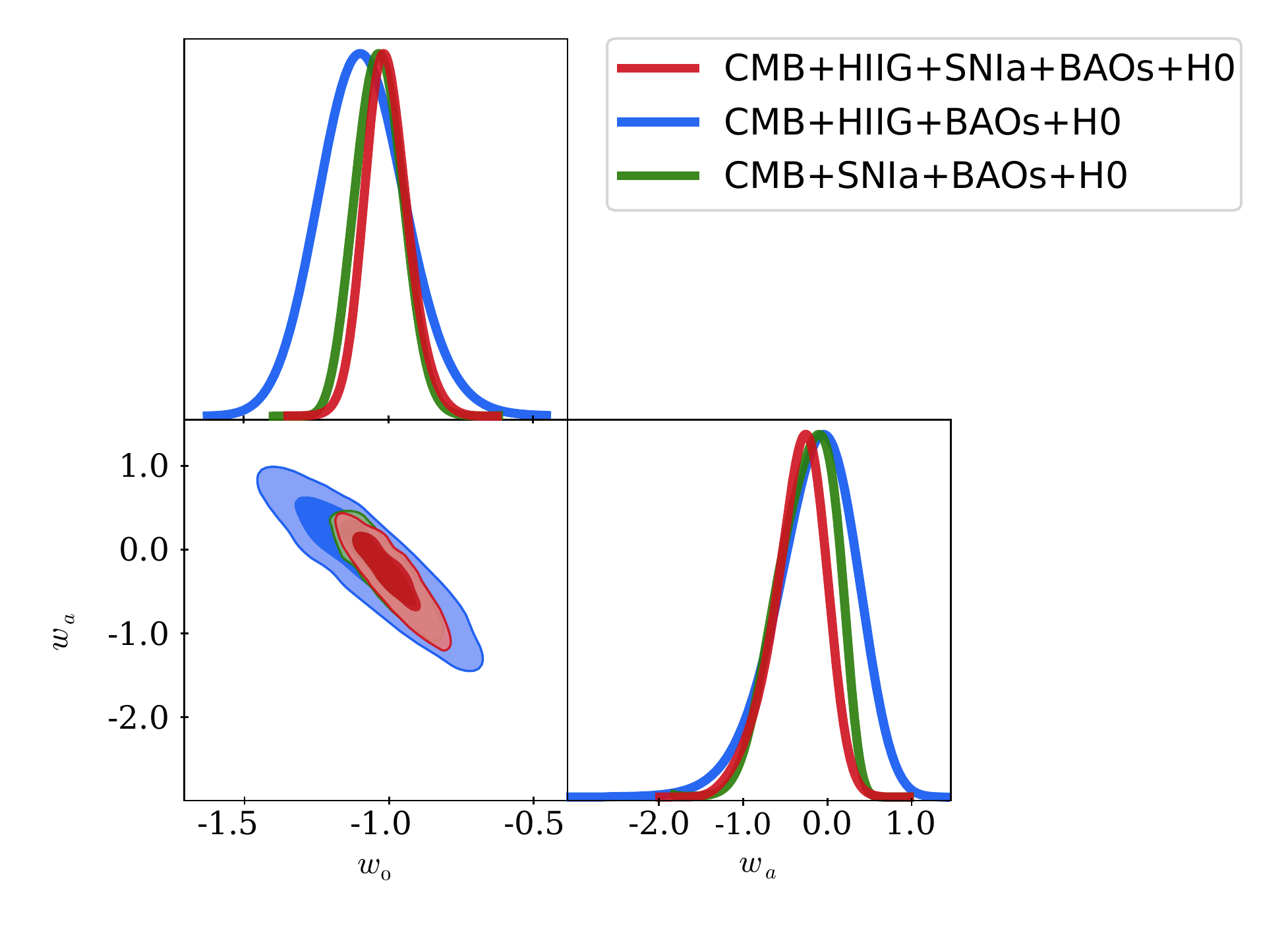}
	\caption{2D contours for the pair $w_0$-$w_a$ from the dataset CMB+HIIG+BAOs+H0 (blue), CMB+SNIa+BAOs+H0 (green), CMB+HIIG+SNIa+BAOs+H0 (red).
        }
	\label{fig:w0wa}
\end{figure}

Our results are listed in table \ref{tab:tabhsw}, while in figure \ref{fig:w0wa} we present the 1 and 2$\sigma$ contours in the $w_0-w_a$ plane when using the combination of HIIG data and SNIa data.
It is worth noting that \lcdm model lies within the 1-$\sigma$ region.

Furthermore, we compare the aforementioned best fit values with those of \citet{Gonzalez-Moran:2021drc}. This work is in alignment with the aforementioned, with the exception that the full Planck spectrum has been utilised in the analysis, in place of the CMB shift parameter, in an effort to include other important cosmological parameters overlooked by the shift parameter such as $\sigma_8$, $\tau_{\text{reio}}$ etc, as well as to have a more complete check of the tested models.

In all cases, \cite{Gonzalez-Moran:2021drc} have found for the combination HIIG+CMB$_{\rm shift}$+BAOs $\Omega_{m,0}=0.298^{+0.018}_{-0.021}$, $H_{0}= 69.8\pm 2.3$km/s/Mpc, $w_0=-1.07^{+0.23}_{-0.32}$ and 
$w_{a}=0.16^{+0.96}_{-0.55}$, while for SNIa+CMB$_{\rm shift}$+BAOs they found $\Omega_{m,0}=0.3011 \pm 0.0085$,
$H_{0}= 68.85\pm 0.99$km/s/Mpc, $w_{0}=-1.062\pm 0.040$ and $w_{a}=0.25^{+0.22}_{-0.19}$.
Evidently, our constraints are compatible within $1\sigma$ with those of \cite{Gonzalez-Moran:2021drc,Gonzalez-Moran:2019uij} and \citet[see figures 4 and 5]{Terlevich:2015toa}.
Finally, our results are also consistent with those of \citet{Scolnic:2017caz}, who found $\Omega_{m,0}=0.300 \pm 0.008$, $H_{0}= 69.057\pm 0.796$km/s/Mpc, $w_{0}=-1.007\pm 0.089$ and $w_{a}=-0.222\pm 0.407$ for SNIa+CMB+BAOs.

%Finally, based on the aforementioned joint statistical analysis we argue that HII galaxies can be used as alternative tracers of the Hubble diagram. THIS HAS BEEN LONG BEEN SHOWN.
Evidently,  we can place constraints on both CPL parameters, marginalizing one over the other, but their corresponding uncertainties remain quite large.
However, what is important to highlight is that the parameter degeneracy in the $w_0-w_a$ plane is large, significantly larger than that of the corresponding SNIa analysis.
Future \hii galaxies data are expected to improve the relevant constraints dramatically, since, based on Monte-Carlo simulations, our team has shown \citep{Chavez16} that a significant improvement is expected when increasing the number of high-z HII galaxies to $\sim 500$, a goal that can be achieved in reasonable observing time with the existing large telescopes, as we have discussed in a number of our previous works.

%\begin{figure}
%	\includegraphics[width=\columnwidth]{omm_w0_all}
%	\caption{2D contour plot of $w_0$ and $\Omega_m$ for the same datasets as \ref{fig:w0wa}.}
%	\label{fig:omm_w0}
%\end{figure}

%	Theoretically the magnitude is calculated through \ref{eq:distmod}, and the luminosity distance in \ref{eq:lumdist} is of course a function of the redshift $z$:
%	
%	\begin{equation}
%		d_L = c(1+z)\int_0^z\frac{d \tilde z}{H(\tilde z)}.
%	\end{equation}
%	
%	The form of the Hubble function $H(z)$ is dependent on the assumed cosmological model.
%	For \lcdm, the late time evolution of the universe is described by:
%	
%	\begin{equation}
%		\label{eq:hubblelcdm}
%		H^2(z) = H^2_0 \left( \Omega_{m0}(1+z)^3 +(1-\Omega_{m0})   \right).
%	\end{equation}
%


%\begin{figure}
%		\includegraphics[width=\columnwidth]{snia}
%		\caption{2D contour plot of $w_0$ and $\Omega_m$ for the CMB+BAO+SNIa dataset.}
%		\label{fig:wowasnia}
%\end{figure}
%
%\begin{figure}
%	\includegraphics[width=\columnwidth]{womm}
%	\caption{2D contour plot of $w_0$ and $\Omega_m$ for the CMB+HIIG+SNIa+BAO+$H_0$ dataset.}
%	\label{fig:ommwboth}
%\end{figure}


\section{Conclusions}
We used \hii galaxies as tracers of the Hubble expansion in a joint analysis, and combined them for the first time with the Planck CMB power spectrum (using CLASS) in order to constrain the parameters for the most popular cosmological models, namely $\Lambda$CDM and CPL.
We also compared the performance of \hii galaxies with that of the widely used Type Ia Supernovae data and found that the best fit parameters of the explored cosmological models are mutually in good agreement. %with small error bars in all cases.
Then we combined HIIG+CMB in a joint statistics with other cosmological probes (SNIa, BAOs, $H_{0}$) 
to place constraints on the CPL cosmological parameters.
We find that the degeneracy of the parameters $w_{0}$ and $w_{a}$ is quite large, which improves however (especially with the reduction of the $w_0$ uncertainty) when confronted with the full set of standard candles (SNIa+HIIG).
This attests to the necessity of further HIIG observations.
Indeed, according to \citet{Plionis_2011} and \citet{Chavez16}, a deviation from $\Lambda$CDM could in principle be detected when using a few hundreds of high redshift HIIG galaxies ($1.5\le z\le 4$).
The current HIIG sample contains in total 181 objects, out of which 74 are high redshift galaxies ($0.6<z<2.6$), which are already yielding very promising results.
Our team, however, is designing the appropriate future KMOS-VLT observations of high-z objects to which  we will add $\sim 100$ additional HIIG, with intermediate redshifts, $z \sim 0.7$, to be observed with MEGARA-GTC as part of the INAOE guaranteed time, as well as to explore the Hubble diagram to redshift $z\sim 3.5$ using MOSFIRE data (Gonz\'{a}lez Mor\'{a}n et al. in preparation).
Hence, we argue that future HIIG data are expected to substantially improve the relevant constraints (especially on $w_{a}$) and thus the validity of dynamical dark energy on extragalactic scales will be effectively tested.
We are also planning to utilize the \hii galaxies Hubble expansion probe to constrain modified gravity models in a future work.

%We find similar results in our analyses, within the \lcdm model, however,
%the HIIG data perform differently in the case of the CPL parameterisation.
%Although the \lcdm model is still contained within the $1\sigma$ CL of both the SNIa and the HIIG analyses, the HIIG favour a larger value for $w_0$ than the SNIa data.
%We expect that future surveys will provide 
%stricter restrains for the parameters.
%It would be interesting to confront more alternative models with both datasets, with the interest in non concordance models increasing.

%%%%%%
\section*{Data Availability}

The main analysis of the present work is based on \hii galaxy data, which have been made available in \cite{Gonzalez-Moran:2021drc}, at https://doi.org/10.1093/mnras/stab1385, \cite{Gonzalez-Moran:2019uij} at https://doi.org/10.1093/mnras/stz1577, \cite{FernandezArenas2018} at https://doi.org/10.1093/mnras/stx2710, \cite{Terlevich:2015toa} at https://doi.org/10.1093/mnras/stv1128\cite{Chavez14} at https://doi.org/10.1093/mnras/stu987.
%%%%%%

\section*{Acknowledgements}
SB acknowledges support by the Research Center for Astronomy of the Academy of Athens in the context of the program ``Tracing the Cosmic Acceleration''.

%%%%%%%%%%%%%%%%%%%%%%%%%%%%%%%%%%%%%%%%%%%%%%%%%%



\bibliographystyle{mnras}
\bibliography{hiig} % if your bibtex file is called example.bib

%\begin{thebibliography}{99}
%\end{thebibliography}

%%%%%%%%%%%%%%%%%%%%%%%%%%%%%%%%%%%%%%%%%%%%%%%%%%

%%%%%%%%%%%%%%%%% APPENDICES %%%%%%%%%%%%%%%%%%%%%

%%%%%%%%%%%%%%%%%%%%%%%%%%%%%%%%%%%%%%%%%%%%%%%%%%

% Don't change these lines
\bsp	% typesetting comment
\label{lastpage}
\end{document}